# Study on State-of-the-art Cloud Services Integration Capabilities with Autonomous Ground Vehicles

Praveen Damacharla
*Dept. of Electrical Engineering and Computer Science*
*The University of Toledo*
Toledo, Ohio, USA
Praveen.Damacharla@rockets.utoledo.edu

Ahmad Y Javaid
*Dept. of Electrical Engineering and Computer Science*
*The University of Toledo*
Toledo, Ohio, USA
Ahmad.Javaid@utoledo.edu

Dhwani Mehta
*Dept. of Electrical Engineering and Computer Science*
*The University of Toledo*
Toledo, Ohio, USA
Dhwani.Mehta@rockets.utoledo.edu

Vijay K. Devabhaktuni
*Dept. of Electrical Engineering and Computer Science*
*The University of Toledo*
Toledo, Ohio, USA
Vijay.Devabhaktuni@utoledo.edu

*Abstract*— Computing and intelligence are substantial requirements for the accurate performance of autonomous ground vehicles (AGVs). In this context, the use of cloud services in addition to onboard computers enhances computing and intelligence capabilities of AGVs. In addition, the vast amount of data processed in a cloud system contributes to overall performance and capabilities of the onboard system. This research study entails a qualitative analysis to gather insights on the applicability of the leading cloud service providers in AGV operations. These services include Google Cloud, Microsoft Azure, Amazon AWS, and IBM Cloud. The study begins with a brief review of AGV technical requirements that are necessary to determine the rationale for identifying the most suitable cloud service. The qualitative analysis studies and addresses the applicability of the cloud service over the proposed generalized AGV's architecture integration, performance, and manageability. Our findings conclude that a generalized AGV architecture can be supported by state-of-the-art cloud service, but there should be a clear line of separation between the primary and secondary computing needs. Moreover, our results show significant lags while using cloud services and preventing their use in real-time AGV operation.

## I. INTRODUCTION

The past two decades of vehicle research is characterized by intense research and development, with the aim of making on-road vehicles smarter, safer, and enjoyable by enhancing the driving experience [1]. Recent ventures and advancement of various technologies have led to the development of autonomous ground vehicles (AGVs). Today, a typical car with several autonomous features is likely to contain an onboard computer, a GPS, a radio transceiver, a short-range collision detection device, cameras, and other sophisticated sensing devices [2]. AGV refers to a vehicle which is capable of path planning and guiding, meaning it is an intelligent system that is able to gather and process information and make decisions that facilitate its movement [3]. Key factors for the success of these technologies lie in the ideology that a mobile robot must have the capacity to localize itself, gather and analyze data from its environment, make appropriate decisions in response to the perceptions, and control actuators to facilitate movement [4]. To achieve these capabilities with the highest accuracy, AGVs need computing power that can be limited by onboard computers. The advancement of wireless networking and the rapidly expanding capabilities of the internet have facilitated increased use of mobile cloud technologies [4]. In the recent past, the advent of Vehicular Ad-Hoc Network (VANET) standards has made information sharing of moving vehicles easier and will be able to bring mobile cloud technology to AGVs in the near future [5].

One prominent example that inspired us is performance enhancement in natural language processors by integration of cloud computing that has facilitated the use of deep learning algorithms, big data processing, and *daily* training. Various organizations have invested and contributed to these systems, including Google, Amazon, and Microsoft. It is imperative to investigate how each off-the-shelf cloud services can contribute to the advancement of AGVs. The intent of this short paper is to present a qualitative and comparative study of commercial cloud service integration capabilities with AGVs.

The objectives of the study include:
- Identify computing requirements of AGVs
- Identify the operating performance of selected cloud services
- Provide a qualitative and computational analysis

The study includes a short literature and requirement review to gather insights from existing research and find AGV computing requirements, a methodology of how the study is accomplished, results with their discussion, and conclusions with suggestions.

## II. LITERATURE REVIEW

The evolution of self-driving cars began with the early driver assistance systems that were based on sensor data processing technique developments [1]. The efficiency of the AGVs has been facilitated by increased R&D over the years. The second generation of vehicles introduced driver assistance systems that use sensors to measure the external state of the vehicle and offer the driver information and warnings that improve the driving experience. These sensors are imperative in the self-driving vehicles, helping in aspects such as vision, LIDAR (Light Detection and Ranging), radar, ultrasonic range (SONAR), GPS (Global Positioning System), and inter-vehicle communication [4]. The basic principle of how AGV systems operate was centered on following a pre-defined route that was monitored by onboard processing devices. The processing mechanism that facilitates real-time data fusion from all these sensors is critical for AGVs. The virtual map offers a modern approach to the guidance systems in which laser sensors and GPRS (General Packet Radio Services) are used to help the AGV to identify its location. By developing the virtual map, the AGV creates the route to follow. The route realization time is short compared to traditional fusion techniques, which were slow and unreliable. A move from the traditional vehicle infrastructure offers a higher level of technology and new features, including safety bumpers, optical scanning instruments, obstacle scanner, and GPS module. The comparison of different factors in a traditional vs recent AGV is presented in Table I. Recent AGVs are dynamic, can detect the path using online information, and contains collision avoidance systems. The vehicles also collect sensor data from neighboring vehicles (V2V) to facilitate effective autonomous driving [6]. The components and requirements of the AGV are presented in Figure 1.

TABLE I. BASIC FACTORS COMPARISON OF TRADITIONAL AND AUTONOMOUS VEHICLE SYSTEMS

| Comparison | Traditional Vehicle | Autonomous Vehicles |
|---|---|---|
| Maneuvering | Movements based on the guiding elements, magnetic tape among others | free movement |
| Integration | Expensive and lengthy | Easy and simple |
| Obstacle Detection | Stop and wait | Move around |
| Collision Warnings | Voice warning and car still moving | Voice warning plus stop car |
| Traffic management | Through GPS | Through GPS+Drones |
| Communication | Through OBUs | Through OBUs+RSUs |

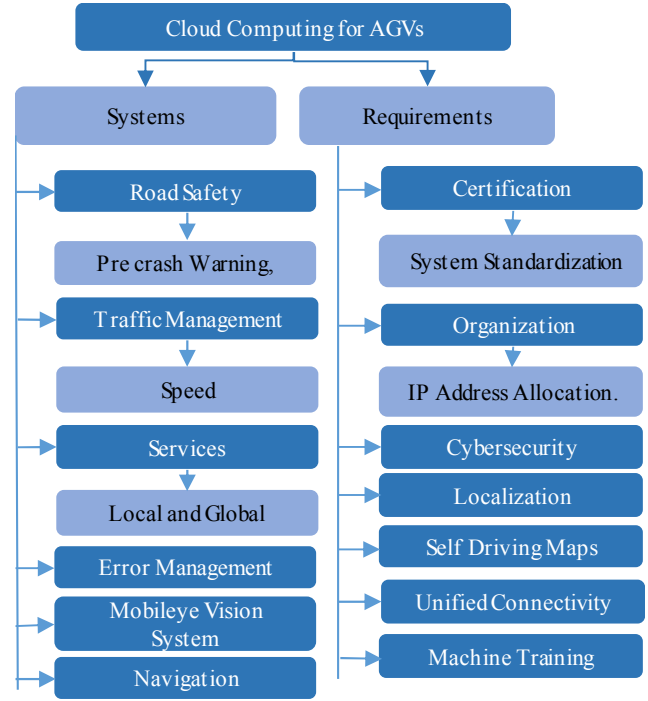

Figure 1. Components and requirements of AGVs

Recently, the private sector, with support from various government agencies, has increased investment in the development of technologies that will support the deployment of AGVs. The technologies include the GPS-based systems that allow the vehicle to gather information from cellular signals and suggest alternative routes in congested roads, helping human drivers make informed decisions [7]. The latest AGVs use GPS navigation modules that can communicate with other devices. These AGV systems consist of the vehicle, onboard control, management, communication, and navigation systems. These components interact through computing and networking systems to support the performance of the vehicle [8]. Some vehicles already have existing autonomous functionalities that include self-parking and collision avoidance features. Fully autonomous vehicles are computer-driven, which makes them rely on data analysis and efficiency of onboard systems. The increased use of cloud computing in robotics offer a new approach to access and process large amounts of data that can enhance the use of AGVs. These insights show that robotics should be capable to solve complex tasks and be enhanced by facilitating regular learning and access to a larger database. The use and applicability of cloud-based services have been coupled with increased research in many domains, including AGVs [4]. The existing frameworks for cloud-based AGV architecture include cloud computing, onboard computing, data storage, and networking [3].

The use of cloud services applies to the existing provisions of intelligent vehicle grids, VANETs, and the vehicular cloud. The architecture is comprised of sensors on the vehicles and roads that offer a large amount of data every second, control units, power drivers, and other elements presented in Figure 1. The vehicular cloud represents the internet of vehicles that entails all protocols and services needed for an AGV to function efficiently and safely [6, 9].

III. METHODOLOGY

The study is accomplished by proposing a generalized cloud architecture for AGVs that is used to perform a thorough application-relevant comparison of cloud service providers using the CloudCmp computational models [10, 11]. We choose the cloud service providers to study and compare based on two criteria: architecture compatibility and computing services available. However, our measurement approach is easily extensible to other providers. Our comparative and qualitative study includes four cloud service providers: Google Cloud [12], Microsoft Azure [13], Amazon AWS [14], and IBM Cloud [15]. Based on selection criteria, these services proved to be compatible with generalized AGV architecture and provide cutting-edge tools that can be used to build next-generation technology with little effort.

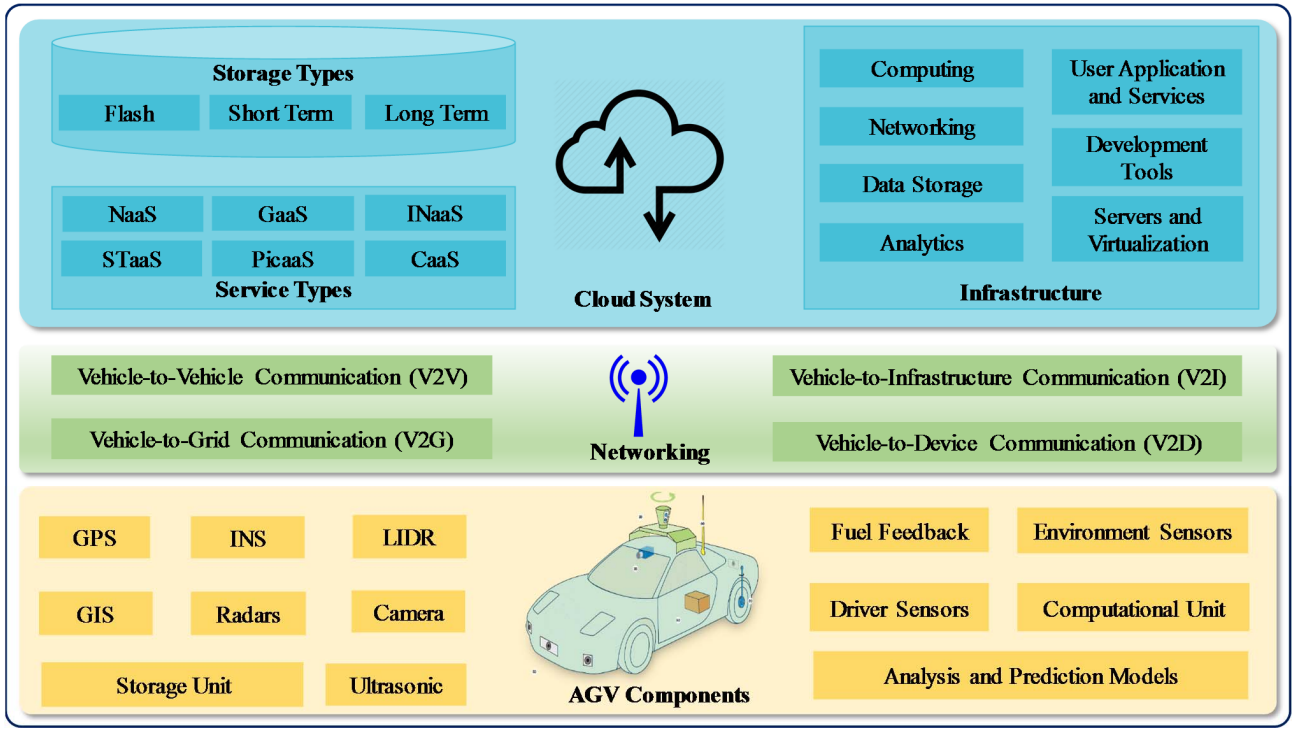

Figure 2. Generalized Cloud Architecture for Autonomous Ground Vehicles

*A. Architecture*

The proposed generalized architecture contains three layers on which the performance of AGV relies. These include the cloud service, the networking/communications, and the AGV, as shown in Figure 2. The proposed architecture assumption is that the OBUs (Onboard Units), maps, localization units, and the navigation hardware units can communicate with the internet via the Wireless Access in Vehicular Environment (WAVE) without interpretation [5, 16].

The first layer has major importance as the data stored in the cloud services can be used to construct next-generation AGVs incorporate perspective. The layer consists of various services like the NaaS (Network as a Service), INaaS (Information as a Service), STaaS (Storage as a Service) [5, 16]. It has two sub-categories, namely storage types, and infrastructure. The information gathered from the inside of the vehicle is stored and processed using one of the three storage types: flash, short-term, or long-term storage. Each storage type varies in the speed of operation and accessibility. The computation is used to build machine learning models, online learning, data preprocessing and post-processing models, and AGV driving profile development.

The second layer is the networking layer, which includes four parts: V2V (Vehicle-to-Vehicle Communication), V2G (Vehicle-to-Grid Communication), V2I (Vehicle-to-Infrastructure Communication), and V2D (Vehicle-to-Device Communication). In V2V abnormal behavior are reported via EWMs (Emergency Warning Messages) to the cloud and the nearby vehicles. The operational data is exchanged between the vehicles and the cloud by V2I over the internet and satellite. The V2G is used for the BEV (electric vehicles) and PHEV (plug-in hybrids) vehicles. The V2D is used for simpler vehicle–driver communications.

The third layer is the AGV component layer, which has the responsibility of gathering the necessary information, such as the behavior/ health of the driver through the use of driver sensors. Other sensors include the INS (Inertial Navigation Sensor), environmental sensors, GPS, GIS (Geographic Information System), camera, and fuel feedback. The driver's intentions and reflexes are predicted, and all the information gathered are stored in the primary vehicle as well as in the cloud. The cloud storage can be further used for real-time application purposes. Proposed architecture operations include the AGV gathering surrounding environment information using sensors and processing them onboard for navigation and communication. The sensor data then goes to the cloud service through the networking layer, where cloud services can use the data for daily training, which in turn can improve and update the AGV models.

*B. Implementation*

This study is accomplished through CloudCmp [10] computation model and quantitative analysis, which consists of all three layers: cloud computing, OBU computing, and networking, as this benchmarking facilitates a comprehensive interpretation and comparison of the data. A comparison is done among four major cloud providers: Microsoft Azure [11], IBM Cloud [12], Google Cloud [13], and Amazon AWS [14]. We have anonymized these four cloud providers randomly and will refer to them as A, B, C, and D (not in any specific order) due to legal concerns of our implementation and results section. The study is accomplished by undertaking rigorous research using data collection at different periods of time over 15 day time intervals to represent compressive time effects on data and cumulative results presented in section IV. By critically examining the data gathered, we are able to formulate a short report on the key objectives of success by different cloud providers based on our generalized architecture presented in this short paper. Table II presents the different cloud instances created.

TABLE II. INFORMATION OF THE CLOUD INSTANCES WITH THEIR LOCATIONS

| Service Provider | Data Center Name | Number of Cores | Price | Location |
|---|---|---|---|---|
| A | A1 | 8 | $0.24/hr | US |
|   | A2 | 4 | $0.12/hr | US |
| B | B1 | 4 | $0.48/hr | Europe |
|   | B2 | 8 | $0.96/hr | US |
| C | C1 | 4 | $0.03/hr | US |
|   | C2 | 4 | $0.06/hr | Europe |
| D | D1 | 4 | $0.48/hr | US |
|   | D2 | 4 | $0.96/hr | US |

This implementation is based on network, storage and computation models. The tools that were used are iperf (network performance measuring tool), ping, a java based client to use the APIs based on the information provided by the implementations referenced by the providers, and SPECjvm2008, respectively. The iperf and ping tools were run between the two instances to measure the latency and throughput (intra-cloud). The sizes of the TCP were controlled to avoid bottleneck since the larger window size does not affect the throughput. To measure the latency, we ping the instances over 250 PlanetLab vantage points. This was done by instantiating the instance provider's data centers. The storage API is used to put and get data from the service. To improve the latency, the client has modifications over the referenced implementations. To avoid SSL overheads, it uses the HTTP persistent connections. The variable data size helps us to concretely understand bottlenecks of throughput and latency of the storage. The number of simultaneous requests has been varied to achieve the best throughput. The experiment was repeated at various times since the performance is impacted by the network as well as by the load on the client.

For computation, Java is used since all four providers support it. The model includes multiple CPU tasks like computations, further extended by I/O and memory tasks. To be consistent with all providers, the tasks are only run for 25 seconds. Tasks are performed on the virtual instances and record their finishing time, while the instances are provided by the clouds. For the instances that have multiple CPU cores, the tasks are run in multiple threads simultaneously to evaluate multi-threading performance. Furthermore, each task's finishing time plus its published service per hour is used to compute the cost.

IV. RESULTS AND DISCUSSION

The services of the cloud providers differ due to a wide range of artificial intelligence and machine-learning capabilities. These services are quickly maturing as the companies seek to meet the emerging customer needs as well as the expanding capability of the cloud-based technologies. To facilitate effective evaluation of the services proposed in this study, a comparison is made to determine their applicability in the vehicular cloud for AGVs. As seen in Table II, the measured instance types are referenced as A, B, C and D. Performance computation calculation is done for measuring the instance types. Instances from Linux are used for the experiment. Figure 3 shows combined finish time of selected tasks based on CPU, memory, and board. Surprisingly, C2 with a relatively lower number of cores and price in comparison to A2, B2 and D2 provide better performance for a combined finish time, while B1 has the lowest combined finish time of all the instances. These bars depict the median of the samples measured. Each task was repeated 30 times per instance for 10 instances (total of 30 samples per task per instance). We conclude that a major difference is seen in the combined performance of various tasks of the four cloud providers. Another observation is that the higher the instances, the shorter the task finishing time. Reasons for including this: Firstly, the instances may have faster CPUs. Secondly, the instance may suffer from higher reserve contention, due to poor multiplexing techniques and high load. Figure 4 depicts the combined TCP throughput for the four cloud providers. It is seen from Figures 3 and 4 that surprisingly, the cloud providers with high instances like B2 do not have high throughput. The throughput of C1 and C2 are almost the same since the number of cores is the same and there is less cost difference in both, while D1 and D2 also have similar results but at a higher price difference. Figure 5 shows the CDF of combined get and put execution response for the various cloud instances. For experimenting reasons, we have repeated each operation multiple times. It is clear from the graph that the services show high range variation in response time. The services provided by provider B is significantly less as compared to other service providers. The response time for D is significantly decreased after the 60$^{th}$ percentile since it does not store the indices over the non-key fields. Providers A and D have a better indexing strategy than the others. A brief survey on security, compliance, archival storage, and manageability of storage is compared in Table III.

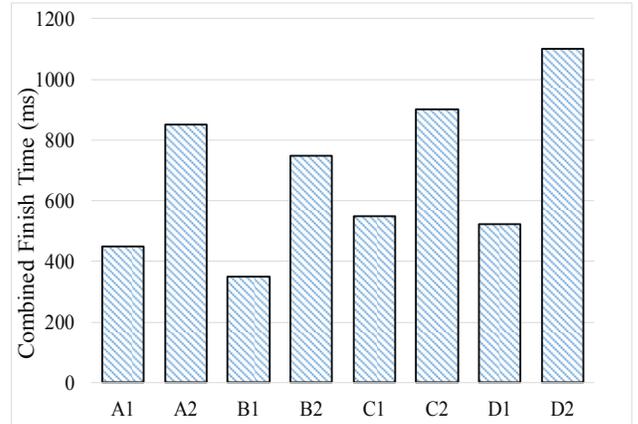

Figure 3. The combined (CPU + Memory + Board) Finish Time of selected tasks for various cloud instances

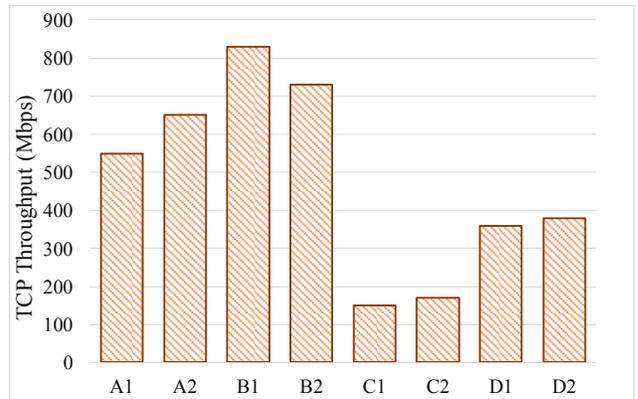

Figure 4. The combined (CPU + Memory + Board) TCP Throughput of selected tasks for various cloud

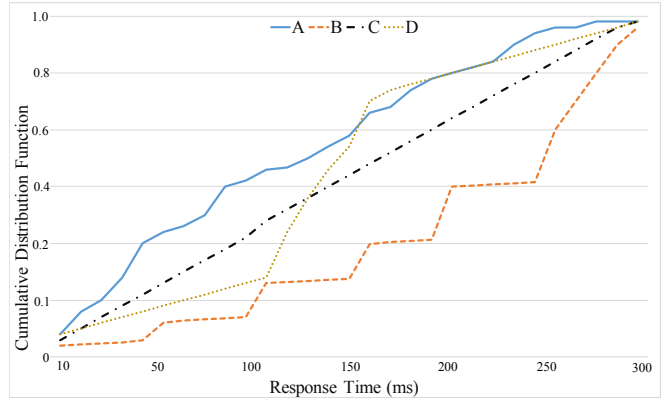

Figure 5. The Cumulative Distribution Function (CDF) of combined Get and Put execution response for various cloud instances

TABLE III: SECURITY, COMPLIANCE AND STORAGE OFFERED BY THE FOUR PROVIDERS.

| Provider Components | Google Cloud | Amazon AWS | Microsoft Azure | IBM |
|---|---|---|---|---|
| Security | encryption by default | Armor, third party involved. | Vormetric Inc, third party involved | Third party involved. |
| Archival Storage | Nearline and Coldline | Glacier | CoolBlob | Object Storage |
| Manageability of storage | Container Engine | EC2 Container Service | Container Services | Container Service |

## V. CONCLUSION AND RECOMMENDATIONS

This study intended to investigate the capability of integration of AGVs with state-of-the-art cloud services. The study offers a review of requirements on the key requirements of AGVs, thus further conducting a critical analysis of the strengths and weaknesses of the four major cloud service providers. We performed systematic procedures that compared the performance of four cloud providers in dimensions such as throughput and task finish time, which are considered of greatest importance to AGV needs. In these virtual instances, there were highly noticeable throughput and time variations among different cloud providers while minor variations in network and storage services were also observed. All of the cloud services showed that due to a task finish time of over 300ms, cloud services could not be used for real-time applications. We conclude that developments have to divide computing tasks as primary and secondary so that cloud services can be used for secondary tasks that can bear the latency involved. Example of primary task obstacle avoidance and examples of secondary tasks include daily training, driving profile storage, and traffic data processing. A cloud service for mobile applications requires the cloud platforms to be customized to meet these needs. In this light, there is a need of intensive cooperation between the AGV developers and the cloud service vendors for successful implementation. There is also a need for intensive research that will help identify the key differences that exist in the cloud platforms regarding the development of AGVs. Our future work includes a comprehensive computational analysis of cloud services using dSPACE GmbH hardware equipment to observe real-time performances at a much deeper level.

ACKNOWLEDGMENTS

Authors appreciate the support of the Paul A. Hotmer Family CSTAR (Cybersecurity and Teaming Research) Lab, EECS (Electrical Engineering and Computer Science) Department at the University of Toledo and Ohio Federal Research Network (OFRN).